# Observation of spin-orbit magnetoresistance in CoFeB/heavy metal/MgO with existence of both spin Hall effect and Edelstein effect

Haoran Ni, Shuangfeng Li, Qihan Zhang, Jiguang Yao, Yongwei Cui, Xiaolong Fan*, Desheng Xue

*Key Laboratory for Magnetism and Magnetic Materials of the Ministry of Education*
*Lanzhou University, Lanzhou, 730000, People's Republic of China*

Corresponding author: Xiaolong Fan
E-mail address: fanxiaolong@lzu.edu.cn

**Abstract**

In this paper, we report the observation of spin-orbit magnetoresistance (SOMR) in ferromagnetic metal/heavy metal/MgO system. We measure the magnetoresistance as the function of the thickness of heavy metal (HM) for CoFeB/HM/MgO and CoFeB/HM films where HM = Pt and Ta. Besides the conventional spin Hall magnetoresistance (SMR) peak, the evidence of the SOMR is indicated by another peak of the MR ratio when the thickness of HM is around 1 ~ 2 nm for CoFeB/HM/MgO films, which is absent for CoFeB/HM films. We speculate the SOMR observed in our experiment originates from the spin-orbit coupling at the HM/MgO interface. We give the boundary conditions of our samples and calculate the theoretical magnetoresistance based on spin diffusion equation. Based on the theoretical results, we can explain the two peaks we observe separately comes from the spin current generated by spin Hall effect and by Edelstein effect.

Magnetoresistance (MR) effect has long been studied, and their tunable property may provide potential application for future memory devices. In recent years, there have been a lot of reports about new unconventional magnetoresistances. The spin Hall magnetoresistance (SMR), which is the result of the interaction between spin Hall effect (SHE) and inverse spin Hall effect (ISHE) [1], depends on the angle between magnetization and spin polarization, and it has been reported in several heterostructure systems[2,3]; The recently reported Hanle Magnetoresistance (HMR) depends on the direction and strength of external magnetic field, and it appears in thin metal film which has a strong spin-orbit coupling[4,5]; The Rashba-Edelstein Magnetoresistance (REMR) is reported in Bi/Ag/CoFeB trilayers, and this magnetoresistance is the result of interfacial spin-orbit coupling and spin-current reflection in the metallic heterostructure[6].Those unconventional MR effect can help us understand the spin-charge conversion by either bulk or interfacial effect better.

Recently, a new type of magnetoresistance was predicted[7], called spin-orbit magnetoresistance (SOMR). It is first proposed in normal metal(NM)/ferromagnetic insulator(FI) bilayers that the Rashba type spin-orbit interaction (SOI) at the interface will give rise to an additional Hamiltonian, and the magnetoresistance related to this will show a maximum as the thickness of the NM layer increases. The first experimental observation of SOMR was done by Lifan Zhou *et al.*[8] By changing the thickness of Pt, the heavy metal layer goes from separate islands to continuous layer, thus changing the strength of spin-orbit interaction at the surface to separate SOMR from SMR. However, the works mentioned above only deal with one interface, where the Rashba SOI and spin absorption and reflection happen at the same place. It is hard to experimentally study individually the properties of SOMR. Therefore, we decide to study the SOMR in FM/HM/MgO trilayers, where the HM/MgO interface is the Rashba interface, and the spin absorption and reflection happen at FM/HM interface. In this way, it will be theoretically easier because we can apply the conventional spin diffusion equation and just adjust the boundary conditions. Experimentally, the SOMR effect can be directly identified by contrasting the magnetoresistance of bilayers and trilayers, and the properties of SOMR can be studied by changing the structure parameters such as the thickness and elements of HM.

We fabricated four sets of samples: (i) CoFeB(5 nm)/Ta($d_{Ta}$)/MgO(3 nm); (ii) CoFeB(5 nm)/Ta($d_{Ta}$); (iii) CoFeB(5 nm)/Pt($d_{Pt}$)/MgO(3 nm); (iv) CoFeB(5 nm)/Pt($d_{Pt}$); All of them were fabricated on Si (100) substrates using magneton sputtering under room temperature, the background pressure was lower than $2 \times 10^{-7}$ Torr, and the pressure of Argon gas was controlled to be 5 mTorr. The thickness of the layers are determined by X-ray reflectivity (XRR), a typical data for CoFeB(5 nm)/Pt(10 nm) is shown in FIG. 1(a). From the number and the intensity of the oscillations we can determine the thickness and roughness of our samples.

It is known that, AMR depends on the angle between magnetization and charge current, SMR and SOMR depends on the angle between magnetization and spin polarization. Since Both SHE and EE would result in spin polarization along $\hat{y}$ axis (i.e. in-plane perpendicular to electrical current), a mixed signal due to SMR and SOMR can be identified from the angular-dependent MR measurements in $\gamma$ plane, which is quite different from the AMR measured in $\beta$ plane, as shown in Fig. 1(b). The definition of the angles is shown in the picture, and the external magnetic field is fixed to be ~ 1.35 $T$. Because those two directions are both out-of-plane directions, the saturated magnetic field under that circumstance will be very high, therefore the original MR plots we get are not ideal sinusoidal curves and the misalignment of **H** and **M** has to be considered. We use the equation derived from static equilibrium condition of magnetization:

$$2H \sin(\beta - \beta_M) + \mu_0 M_{eff} \sin(2\beta_M) = 0, \tag{1}$$

to determine the angle of **M**, where $\mu_0 M_{eff}$ is the effective demagnetization field which is determined using Magneto-optical Kerr effect (not shown here).

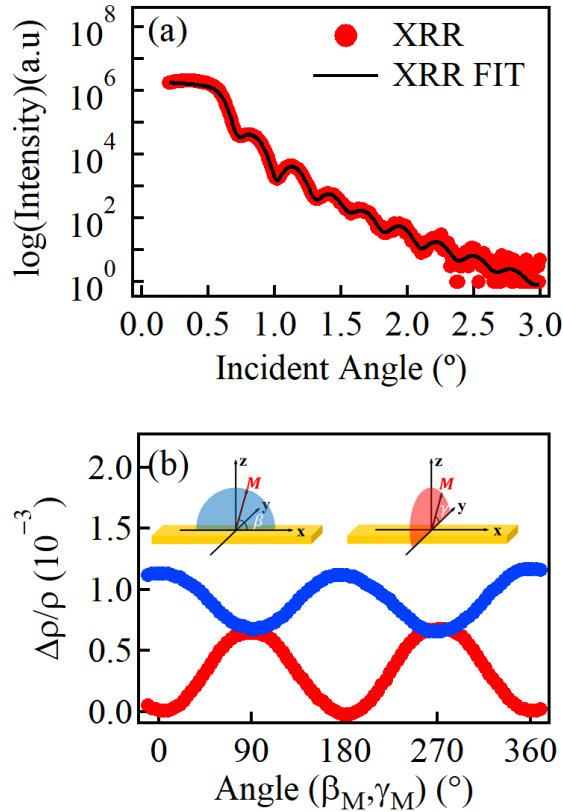

FIG. 1. (a) XRR measurements of CoFeB(5 nm)/Pt(10 nm). The thickness of Pt layer is fit to be 10.2 nm. (b) MR measurements. The red line (SMR) corresponds to $\gamma$ measurement, and the blue line (AMR) corresponds to $\beta$ measurement.

In the paper of Lifan Zhou et al.'s work[8], they discovered that SOMR appears as another peak before the peak of SMR. We observe similar effect in our measurements, as shown in FIG. 2(a) and (c). When the HM layer is thin, the MR effect is very small. As the thickness increases, 2 nm for CoFeB(10 nm)/Ta($d_{Ta}$)/MgO(3 nm) and 1.25 nm for CoFeB(10 nm)/Pt($d_{Pt}$)/MgO(3 nm), the MR effect reaches a maximum, which may be attributed to SOMR effect. Then it decreases for Ta = 3 nm and Pt = 2 nm. The MR effect reaches a maximum again for Ta = 5 nm and Pt = 2.5 nm, which is the sign of the conventional SMR effect. When Ta is thicker than 5 nm and Pt is thicker than 2.5 nm, the MR ratio decreases again.

In order to make sure that the MR we observe is SOMR and that it is generated at the interface between HM(Pt, Ta) and MgO, we did exact the same measurements on CoFeB(10 nm)/Ta($d_{Ta}$) and CoFeB(10 nm)/Pt($d_{Pt}$). The $\Delta\rho/\rho$ diagrams of those two series are shown in FIG. 2(b) and FIG. 2(d). As we expected, the SOMR peak disappears in this case. This result confirms that the MR we observe is SOMR and it originates from the interface between HM layer and MgO layer.

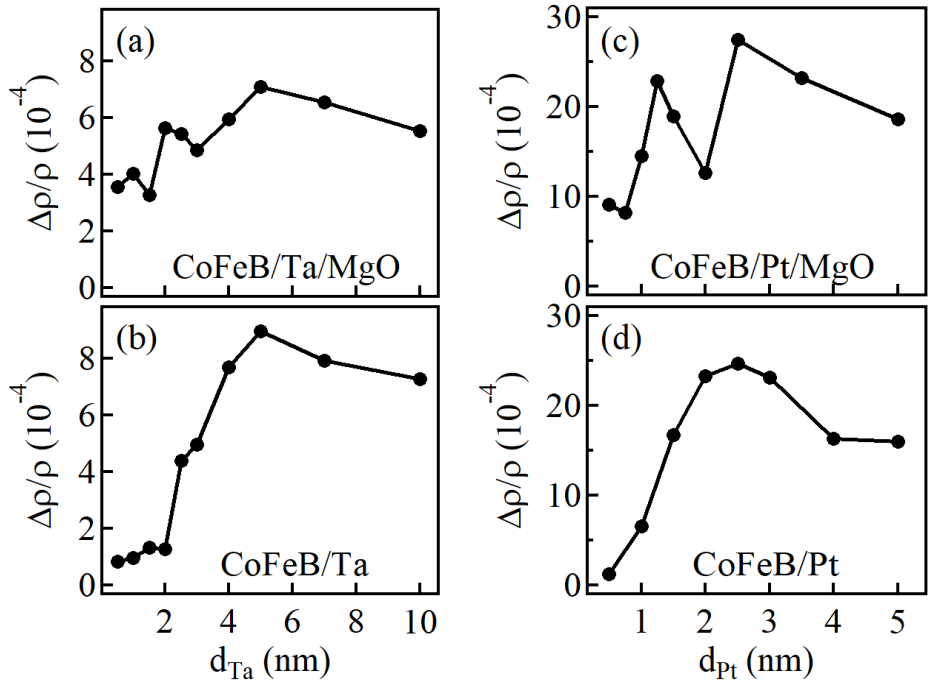

FIG. 2. (a) Ta layer thickness dependence of MR ratio for CoFeB(5 nm)/Ta($d_{Ta}$)/MgO(3 nm). (b) Pt layer thickness dependence of MR ratio for CoFeB(5 nm)/Pt($d_{Pt}$)/MgO(3 nm). (c) Ta thickness dependence of MR ratio for CoFeB(5 nm)/Ta($d_{Ta}$). (d) Pt thickness dependence of MR ratio for CoFeB(5 nm)/Pt($d_{Pt}$).

In our samples, there are two main effects that generate spin current, SHE in HM layer and EE at HM/MgO interface. We give the boundary conditions:

$$z = 0: \quad e\boldsymbol{j}_s^z(z=0) = G_r \hat{\boldsymbol{m}} \times [\hat{\boldsymbol{m}} \times \boldsymbol{\mu}_s(0)] + G_i \hat{\boldsymbol{m}} \times \boldsymbol{\mu}_s(0), \tag{2}$$

$$z = d: \quad \boldsymbol{\mu}_s(d) = (\mu_s^R - \mu_s^0)\hat{\boldsymbol{y}}, \tag{3}$$

where $G_{\uparrow\downarrow} = G_r + iG_i$ is the spin-mixing conductance; $\boldsymbol{\mu}_s(z)$ is the spin accumulation; $\mu_s^R = \eta \frac{3m_e^*}{2e\hbar\epsilon_F}\frac{1}{D(\epsilon_F)}\alpha_{R\_eff}j_c^0$, where $\alpha_{R\_eff}$ is the Rashba parameter, $D(\epsilon_F)$ is the density of states at fermi level; $\mu_s^0 = \frac{2e\lambda}{\sigma}\tanh\frac{d}{2\lambda}j_{s0}^{SH}$, where $\lambda$ is the spin diffusion length in HM, $j_{s0}^{SH} = \theta_{SH}j_c^0$ is the spin current caused by SHE, $\theta_{SH}$ is the spin Hall angle.

The overall spin current in our sample can be written as:

$$\boldsymbol{j}_s^z(z) = -\frac{\sigma}{2e}\partial_z \boldsymbol{\mu}_s - j_{s0}^{SH}\hat{\boldsymbol{y}}, \tag{4}$$

Using spin diffusion equation, we get

$$\boldsymbol{\mu}_s(z) = \mu_s^R \hat{\boldsymbol{y}}\frac{\cosh\frac{z}{\lambda}}{\cosh\frac{d}{\lambda}} + \mu_s^0 \hat{\boldsymbol{y}}\frac{\sinh\frac{d-2z}{2\lambda}}{\sinh\frac{d}{2\lambda}} + \frac{2e\lambda}{\sigma}\frac{\sinh\frac{d-z}{\lambda}}{\cosh\frac{d}{\lambda}}\boldsymbol{j}_s^z(0), \tag{5}$$

Solve $\boldsymbol{j}_s^z(0)$ to obtain

$$\boldsymbol{\mu}_s(z) = \mu_s^R \hat{\boldsymbol{y}}\frac{\cosh\frac{z}{\lambda}}{\cosh\frac{d}{\lambda}} + \frac{2\lambda}{\sigma}\mu_s^R \hat{\boldsymbol{m}} \times (\hat{\boldsymbol{m}} \times \hat{\boldsymbol{y}})Re\frac{G_{\uparrow\downarrow}}{1 + \frac{2\lambda}{\sigma}G_{\uparrow\downarrow}\tanh\frac{d}{\lambda}}\frac{\sinh\frac{d-z}{\lambda}}{\cosh^2\frac{d}{\lambda}}$$

$$+ \frac{2\lambda}{\sigma}\mu_s^R \hat{\boldsymbol{m}} \times \hat{\boldsymbol{y}}\, Im\frac{G_{\uparrow\downarrow}}{1 + \frac{2\lambda}{\sigma}G_{\uparrow\downarrow}\tanh\frac{d}{\lambda}}\frac{\sinh\frac{d-z}{\lambda}}{\cosh^2\frac{d}{\lambda}} + \mu_s^0 \hat{\boldsymbol{y}}\frac{\sinh\frac{d-2z}{2\lambda}}{\sinh\frac{d}{\lambda}}$$

$$+ \frac{2\lambda}{\sigma}\mu_s^0 \hat{\boldsymbol{m}} \times (\hat{\boldsymbol{m}} \times \hat{\boldsymbol{y}})Re\frac{G_{\uparrow\downarrow}}{1 + \frac{2\lambda}{\sigma}G_{\uparrow\downarrow}\tanh\frac{d}{\lambda}}\frac{\sinh\frac{d-z}{\lambda}}{\cosh\frac{d}{\lambda}}$$

$$+ \frac{2\lambda}{\sigma}\mu_s^0 \hat{\boldsymbol{m}} \times \hat{\boldsymbol{y}}\, Im\frac{G_{\uparrow\downarrow}}{1 + \frac{2\lambda}{\sigma}G_{\uparrow\downarrow}\tanh\frac{d}{\lambda}}\frac{\sinh\frac{d-z}{\lambda}}{\cosh\frac{d}{\lambda}}, \tag{6}$$

The first three items of the above function are exactly the same as the $\boldsymbol{\mu}_s(z)$ given in [9], wherein only EE is involved; the left two items characterize the affect brought by SHE. Based on Eq. (4), the spin current is given by

$$\boldsymbol{j}_s^z(z) = -\frac{\sigma}{2e\lambda}\mu_s^R\hat{\boldsymbol{y}}\frac{\sinh\frac{z}{\lambda}}{\cosh\frac{d}{\lambda}} + \frac{1}{e}\mu_s^R\hat{\boldsymbol{m}}\times(\hat{\boldsymbol{m}}\times\hat{\boldsymbol{y}})Re\frac{G_{\uparrow\downarrow}}{1+\frac{2\lambda}{\sigma}G_{\uparrow\downarrow}\tanh\frac{d}{\lambda}}\frac{\cosh\frac{d-z}{\lambda}}{\cosh^2\frac{d}{\lambda}}$$

$$+\frac{1}{e}\mu_s^R\hat{\boldsymbol{m}}\times\hat{\boldsymbol{y}}Im\frac{G_{\uparrow\downarrow}}{1+\frac{2\lambda}{\sigma}G_{\uparrow\downarrow}\tanh\frac{d}{\lambda}}\frac{\cosh\frac{d-z}{\lambda}}{\cosh^2\frac{d}{\lambda}} + \frac{\sigma}{2e\lambda}\mu_s^0\hat{\boldsymbol{y}}\frac{\cosh\frac{d-2z}{2\lambda}}{\sinh\frac{d}{2\lambda}}$$

$$+\frac{1}{e}\mu_s^0\hat{\boldsymbol{m}}\times(\hat{\boldsymbol{m}}\times\hat{\boldsymbol{y}})Re\frac{G_{\uparrow\downarrow}}{1+\frac{2\lambda}{\sigma}G_{\uparrow\downarrow}\tanh\frac{d}{\lambda}}\frac{\cosh\frac{d-z}{\lambda}}{\cosh\frac{d}{\lambda}}$$

$$+\frac{1}{e}\mu_s^0\hat{\boldsymbol{m}}\times\hat{\boldsymbol{y}}Im\frac{G_{\uparrow\downarrow}}{1+\frac{2\lambda}{\sigma}G_{\uparrow\downarrow}\tanh\frac{d}{\lambda}}\frac{\cosh\frac{d-z}{\lambda}}{\cosh\frac{d}{\lambda}} - j_{s0}^{SH}\hat{\boldsymbol{y}},$$

Knowing $\boldsymbol{j}_s^z(z)$, we can obtain the charge current induced by spin current in $\hat{\boldsymbol{x}}$ direction $\delta j_c^{ISHE}$

$$\delta j_c^{ISHE} = -\frac{\theta_{SH}\sigma}{2e\lambda}\mu_s^R\frac{\sinh\frac{z}{\lambda}}{\cosh\frac{d}{\lambda}} + \frac{\theta_{SH}}{e}\mu_s^R(m_y^2-1)Re\frac{G_{\uparrow\downarrow}}{1+\frac{2\lambda}{\sigma}G_{\uparrow\downarrow}\tanh\frac{d}{\lambda}}\frac{\cosh\frac{d-z}{\lambda}}{\cosh^2\frac{d}{\lambda}}$$

$$+\frac{\theta_{SH}\sigma}{2e\lambda}\mu_s^0\frac{\cosh\frac{d-2z}{2\lambda}}{\sinh\frac{d}{2\lambda}} + \frac{\theta_{SH}}{e}\mu_s^0(m_y^2-1)Re\frac{G_{\uparrow\downarrow}}{1+\frac{2\lambda}{\sigma}G_{\uparrow\downarrow}\tanh\frac{d}{\lambda}}\frac{\cosh\frac{d-z}{\lambda}}{\cosh\frac{d}{\lambda}},$$

The inverse Edelstein effect (IEE) at HM/MgO interface also contributes to the charge current in $\hat{\boldsymbol{x}}$ direction

$$\delta j_c^{IEE} = \lambda_{IEE}[\boldsymbol{j}_s^z(d)\times\hat{\boldsymbol{z}}]_x$$

$$= -\frac{\lambda_{IEE}\sigma}{2e\lambda}\mu_s^R\frac{\sinh\frac{d}{\lambda}}{\cosh\frac{d}{\lambda}} + \frac{\lambda_{IEE}}{e}\frac{\mu_s^R}{\cosh^2\frac{d}{\lambda}}(m_y^2-1)Re\frac{G_{\uparrow\downarrow}}{1+\frac{2\lambda}{\sigma}G_{\uparrow\downarrow}\tanh\frac{d}{\lambda}} + \frac{\lambda_{IEE}\sigma}{2e\lambda}\mu_s^0\frac{\cosh\frac{d}{2\lambda}}{\sinh\frac{d}{2\lambda}}$$

$$+\frac{\lambda_{IEE}}{e}\frac{\mu_s^0}{\cosh\frac{d}{\lambda}}(m_y^2-1)Re\frac{G_{\uparrow\downarrow}}{1+\frac{2\lambda}{\sigma}G_{\uparrow\downarrow}\tanh\frac{d}{\lambda}} - \lambda_{IEE}j_{s0}^{SH}, \qquad (8)$$

Therefore we have

$$j_{cx} = j_c^0 + \delta j_c^{ISHE} + \delta j_c^{IEE}, \qquad (9)$$

$$\rho_{HM} = \rho_0 + \Delta\rho_1 + \Delta\rho_2(1-m_y^2), \qquad (10)$$

and

$$\frac{\Delta\rho_2}{\rho_{total}} = \zeta\left[\frac{\theta_{SH}}{e}\xi_{EE}\frac{\lambda\tanh\frac{d}{\lambda}}{d\cosh\frac{d}{\lambda}} + \frac{\lambda_{IEE}}{e}\frac{\xi_{EE}}{\cosh^2\frac{d}{\lambda}} + 2\lambda\theta_{SH}^2\frac{\lambda}{d}\tanh\frac{d}{\lambda}\tanh\frac{d}{2\lambda}\right.$$
$$\left.+ 2\lambda\theta_{SH}\lambda_{IEE}\frac{\tanh\frac{d}{2\lambda}}{\cosh\frac{d}{\lambda}}\right]\frac{G_r}{\sigma + 2\lambda G_r\tanh\frac{d}{\lambda}}, \quad (11)$$

within which we let $\mu_s^R = E_x\xi_{EE}$, $\xi_{EE} = \eta\frac{3m_e^*}{2e\hbar\epsilon_F}\frac{1}{D(\epsilon_F)}\alpha_{R\_eff}\sigma$, $G_r \gg G_i$, and consider the current shunting factor

$$\zeta = \frac{1}{1 + \frac{\rho_{HM}d_{FM}}{\rho_{FM}d_{HM}}}, \quad (12)$$

We draw the separate plot of each item of Eq. (11) in FIG. 3(c).

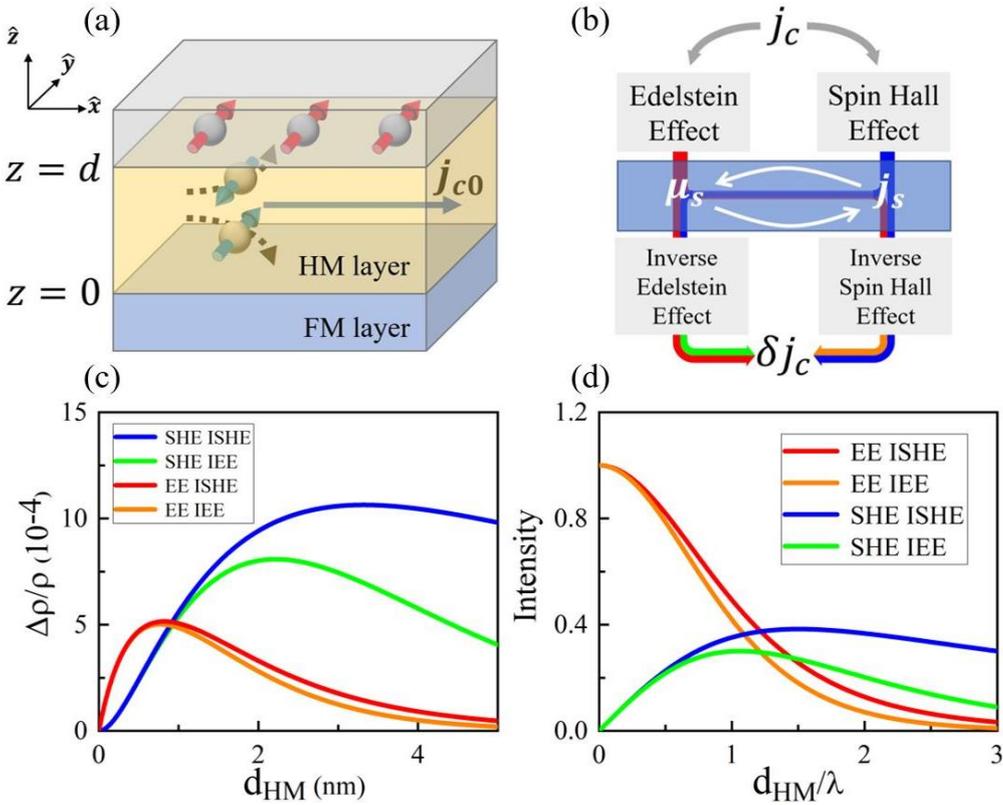

FIG. 3. (a) An illustration of SHE and Edelstein effect (EE) in our sample. Spins with red arrows come from EE, spins with blue arrows come from SHE. (b) Diagram of how different effects influence charge current. Colors of different route correspond to the colors in (c) and (d). (c) Separate plot of each part in Eq. (11). Namely, Blue line: $2\zeta\lambda\theta_{SH}^2\frac{\lambda}{d}\tanh\frac{d}{\lambda}\tanh\frac{d}{2\lambda}\frac{G_r}{\sigma+2\lambda G_r\tanh\frac{d}{\lambda}}$; Green line: $2\zeta\lambda\theta_{SH}\lambda_{IEE}\frac{\tanh\frac{d}{2\lambda}}{\cosh\frac{d}{\lambda}}\frac{G_r}{\sigma+2\lambda G_r\tanh\frac{d}{\lambda}}$; Red line: $\zeta\frac{\theta_{SH}}{e}\xi_{EE}\frac{\lambda\tanh\frac{d}{\lambda}}{d\cosh\frac{d}{\lambda}}\frac{G_r}{\sigma+2\lambda G_r\tanh\frac{d}{\lambda}}$; Orange line:

$\zeta \frac{\lambda_{IEE}}{e} \frac{\xi_{EE}}{\cosh^2\frac{d}{\lambda}} \frac{G_r}{\sigma+2\lambda G_r \tanh\frac{d}{\lambda}}$; The parameters are chosen to be $\theta_{SH} = 0.08$, $\lambda_{IEE} = 0.08$, $\frac{\xi_{EE}}{e} = 0.08\ nm$, $\lambda = 2\ nm$, $G_r = 3 \times 10^{15}\ \Omega^{-1}m^{-2}$, $\sigma = 1 \times 10^7\ \Omega^{-1}m^{-1}$, $\frac{\rho_{HM}d_{FM}}{\rho_{FM}} = 1\ nm$. (d) A more careful study of Eq. (11). To be accurate, Red line: $\frac{\theta_{SH}}{e}\xi_{EE}\frac{\lambda \tanh\frac{d}{\lambda}}{d \cosh\frac{d}{\lambda}}$; Orange line: $\frac{\lambda_{IEE}}{e}\frac{\xi_{EE}}{\cosh^2\frac{d}{\lambda}}$; Blue line: $2\lambda\theta_{SH}^2 \frac{\lambda}{d} \tanh\frac{d}{\lambda} \tanh\frac{d}{2\lambda}$; Green line: $2\lambda\theta_{SH}\lambda_{IEE}\frac{\tanh\frac{d}{2\lambda}}{\cosh\frac{d}{\lambda}}$. For convenience, we set all parameter to be 1 here in order to explicitly see the trend of each curve.

We can see that the spin current generated by EE converts to charge current through ISHE and IEE, and the magnetoresistance of those two reaches a maximum at close HM thickness. The same is true for the spin current generated by SHE, which converts to charge current through ISHE and IEE. Therefore we can explain the two peaks in our experiment in this way. The first and smaller peak comes from the spin current generated by EE and the second peak comes from the SHE. The relative magnitude of those two peaks characterize the relative "amount" of spin current generated by EE and SHE.

By looking more carefully at each function in Eq. (11), we can explain why the peak of SOMR appears at the thickness thinner than that of the peak of SMR. As illustrated in FIG. 3(d), the red and orange curve all relate to EE, which decrease monotonically with increasing thickness. It is reasonable because EE just introduces a spin accumulation at the interface independent of the thickness of HM layer, so the effect will naturally decrease when the thickness of HM increases. The blue and green curve are peak curves. Because SHE is a bulk effect, the thicker the HM layer, the stronger the SHE. However, spin diffusion will be inhibited with thicker HM layer. These two processes will reach a maximum at some point, which gives us peak curves. Besides the four functions we mention in this paragraph, the functions, namely the current shunting factor and $\frac{G_r}{\sigma+2\lambda G_r \tanh\frac{d}{\lambda}}$, owned by every part multiply to be a peak function. Now it is natural that SOMR appears at a thickness thinner than that of SMR. Because the MR function related to EE decreases monotonically with increasing thickness, the peak of it will naturally appear at a thinner HM thickness when multiplied by a peak function.

However, our function fails to explain the two peak phenomenon when we add the four items together. Our experiment result, as well as that in [8], is too sharp to fit. The peak function given by solving spin diffusion equation is too smooth that when we add functions with different peak position together, the difference of peaks cancels out and only gives one peak. Some more precise theories are needed.

**Conclusions**

We report the SOMR effect in CoFeB/Ta/MgO and CoFeB/Pt/MgO system, and confirm that SOMR originates from the interface between HM and MgO. By proposing the boundary condition at HM/MgO interface and solve the spin diffusion equation, we can explain the origin of the two peaks we observe using the function we get. The theory analysis is not satisfactory for the fact that our function cannot fit the data. So we hope that some more accurate theory can be proposed to figure out the mechanism when SHE and EE both exist.

**Acknowledgments**   This project was supported by NSFC (Nos 11674142, 51771099, 11429401 and 51471081), the Program for Changjiang Scholars and Innovative Research Team in University (No IRT-16R35) and the Cuiying Foundation in Lanzhou University.